\documentclass[conference]{IEEEtran}
\IEEEoverridecommandlockouts

\usepackage[utf8]{inputenc}
\usepackage[T1]{fontenc}
\usepackage{cite}
\usepackage{amsmath,amssymb,amsfonts}
\usepackage{graphicx}
\usepackage{textcomp}
\usepackage{xcolor}
\usepackage{bm}
\usepackage{array}
\usepackage{multirow}
\usepackage{booktabs}

\begin{document}
\flushbottom

%-------------------------------------------------------------------------
%  TITLE  (no math/symbols)
%-------------------------------------------------------------------------
\title{Low-Complexity Recurrent Neural Network Detector for
Faster-than-Nyquist Signaling}

\author{\IEEEauthorblockN{Nurettin Safak, Osman Tokluoglu, Enver Cavus}
\IEEEauthorblockA{Department of Electrical and Electronics Engineering,
Ankara Yildirim Beyazit University, Ankara, Turkiye\\
nsafaked@gmail.com, otokluoglu@aybu.edu.tr, ecavus@aybu.edu.tr}}

\maketitle
\IEEEpubid{\makebox[\textwidth][l]{\footnotesize\textbf{979-8-3195-4709-5/26/\$31.00 \copyright\,2026 IEEE}}}

%-------------------------------------------------------------------------
%  ABSTRACT + KEYWORDS  (no math/symbols)
%-------------------------------------------------------------------------
\begin{abstract}
This study proposes a low-complexity, bidirectional, single-pass Elman
recurrent neural network detector for binary phase-shift keying signals
transmitted with faster-than-Nyquist signaling. Since the faster-than-Nyquist
intersymbol interference has a short, finite memory, the classical Elman
recurrent neural network, which contains no gating mechanism, is a sufficient
and parameter-efficient model. The proposed detector processes the received
sequence in both forward and backward directions in a single pass, forming a
learned counterpart of the optimal BCJR forward-backward recursion. Under a
root-raised-cosine pulse over an additive white Gaussian noise channel,
simulations for two compression factors show that the proposed detector, with
only twenty-five to sixty-five trainable parameters, attains a bit error rate
very close to that of the M-BCJR algorithm, while reducing the look-up-table
hardware cost by thirty-eight to forty-six percent and using no explicit
division or exponential operations. A more compact configuration offers up to a
sixty-seven percent reduction at a small performance penalty. To the best of
our knowledge, this is the first study to investigate the classical Elman
recurrent neural network architecture for the faster-than-Nyquist detection
problem.
\end{abstract}


%-------------------------------------------------------------------------
%  I. INTRODUCTION
%-------------------------------------------------------------------------
\section{Introduction}
\IEEEpubidadjcol

Growing data-rate demands and limited spectrum resources have increased the
need for transmission schemes that improve spectral efficiency.
Faster-than-Nyquist (FTN) signaling is an important technique that increases the
data rate by operating beyond the Nyquist limit, introducing a controlled amount
of intersymbol interference (ISI)~\cite{mazo}. However, since FTN inevitably
produces strong ISI, it leads to more complex detection problems at the
receiver. Although optimal algorithms such as BCJR provide high detection
performance, their computational complexity is quite high~\cite{bcjr}.
Therefore, developing detection methods that reduce computational complexity
while preserving high performance has become an important research area.

Various low-complexity detection methods have been proposed to mitigate the ISI
effect in FTN systems. Prlja and Anderson presented a reduced-complexity M-BCJR
algorithm for turbo equalization, achieving near-optimal BER performance over
AWGN channels~\cite{prlja}. Sugiura proposed a frequency-domain equalization
(FDE)-based receiver structure to reduce the time-domain computational
burden~\cite{sugiura}. Bedeer et al.\ realized low-complexity FTN detection using
a successive symbol-by-symbol sequence estimator (SSSSE) with a go-back-K
strategy~\cite{bedeer}. In addition, a reduced-complexity M-BCJR algorithm based
on the Ungerboeck observation model has been proposed, reporting significant
performance gains for $\tau = 0.8$~\cite{ungerboeck}.

Deep-learning-based methods have also attracted growing interest, often as an
auxiliary tool supporting conventional detection algorithms. For example,
LSTM-based RNN architectures have achieved performance close to BCJR in FTN
signal detection~\cite{baek2021}, and SIC-aided structures, sliding-window
methods, and deep-learning-aided list sphere decoding have been
proposed~\cite{song,pan,abbasi}. Machine-learning-based detection in multipath
channels~\cite{baek2024} and deep-learning-aided sum-product detection on factor
graphs~\cite{liu} have also been investigated, while GRU-based architectures
provide BER performance close to the optimal BCJR algorithm for $\tau \geq
0.7$~\cite{gru}. Nevertheless, relatively few studies use deep-learning
architectures directly as standalone FTN detectors. In this context, CNN-based
detectors have been proposed with successful ISI
mitigation~\cite{defilippo,domaincnn}; in particular, a fixed-kernel CNN exploits
the ISI structure of FTN to achieve high performance with lower computational
complexity~\cite{cnn}. More recently, a position-aware attention-enhanced Bi-GRU
detector was proposed to exploit structured ISI information in FTN
signaling~\cite{bigru}. In a similar direction, a self-attention
transformer-based detector~\cite{transformerftn} and a Kolmogorov--Arnold
network (KAN)-based detector~\cite{kanmlp} have also been investigated as
standalone FTN detectors.

In this study, a low-complexity Elman RNN-based detector is proposed for FTN
detection. While more complex gated architectures (LSTM, GRU) dominate the FTN
literature, simpler recurrent structures have been overlooked. The classical
Elman RNN~\cite{elman} processes the received sequence bidirectionally in a
single pass, forming a learned counterpart of the BCJR forward-backward
recursion. To the best of our knowledge, this is the first study to investigate
the Elman RNN for FTN detection.

%-------------------------------------------------------------------------
%  II. SYSTEM MODEL
%-------------------------------------------------------------------------
\section{System Model}

For a band-limited baseband pulse shape $g(t)$, the transmitted FTN signal is
expressed as
\begin{equation}
s(t) = \sum_{k} a_k\, g(t - k\tau T).
\end{equation}
Here, $a_k$ denotes the $k$-th modulation symbol (BPSK in this study), $T$ the
Nyquist symbol interval, and $\tau \in (0,1)$ the compression factor. Reducing
$\tau$ places the symbols more closely together, thereby increasing spectral
efficiency. However, this causes intersymbol interference (ISI) due to pulse
overlap. Therefore, the ISI effect must be taken into account for accurate
symbol estimation at the receiver.

In this study, a root-raised-cosine (RRC) pulse with roll-off factor $\beta =
0.35$ is used for the transmit and receive filters. The pulse energy is
normalized as
\begin{equation}
\int_{-\infty}^{+\infty} |g(t)|^2 \, dt = 1.
\end{equation}
In FTN transmission, the spectral efficiency increases approximately by a factor
of $1/\tau$. For example, for $\tau = 0.9$ a spectral efficiency gain of about
11\% is obtained.

Transmission takes place over an additive white Gaussian noise (AWGN) channel.
The sampled signal obtained after matched filtering at the receiver is expressed
as
\begin{equation}
y(n\tau T) = \sum_{k} a_k\, x\bigl((n-k)\tau T\bigr) + w(n\tau T)
\end{equation}
\begin{equation}
x(t) = g(t) * g(-t).
\label{eq:mf}
\end{equation}
Equation~\eqref{eq:mf} denotes the matched-filter response and $w(n\tau T)$ the
noise at the sampling instant. Since the FTN sampling interval $\tau T$ is
smaller than the Nyquist interval, the noise at the matched-filter output is not
white; it is colored noise whose autocorrelation is proportional to the ISI
response $\mathbf{X}$. While the reference M-BCJR algorithm accounts for this
colored noise in its branch metric, the proposed RNN learns this statistic
directly from data, avoiding any explicit estimation or inversion of the noise
covariance at the receiver.

\begin{figure}[t]
\centering
\includegraphics[width=0.99\columnwidth]{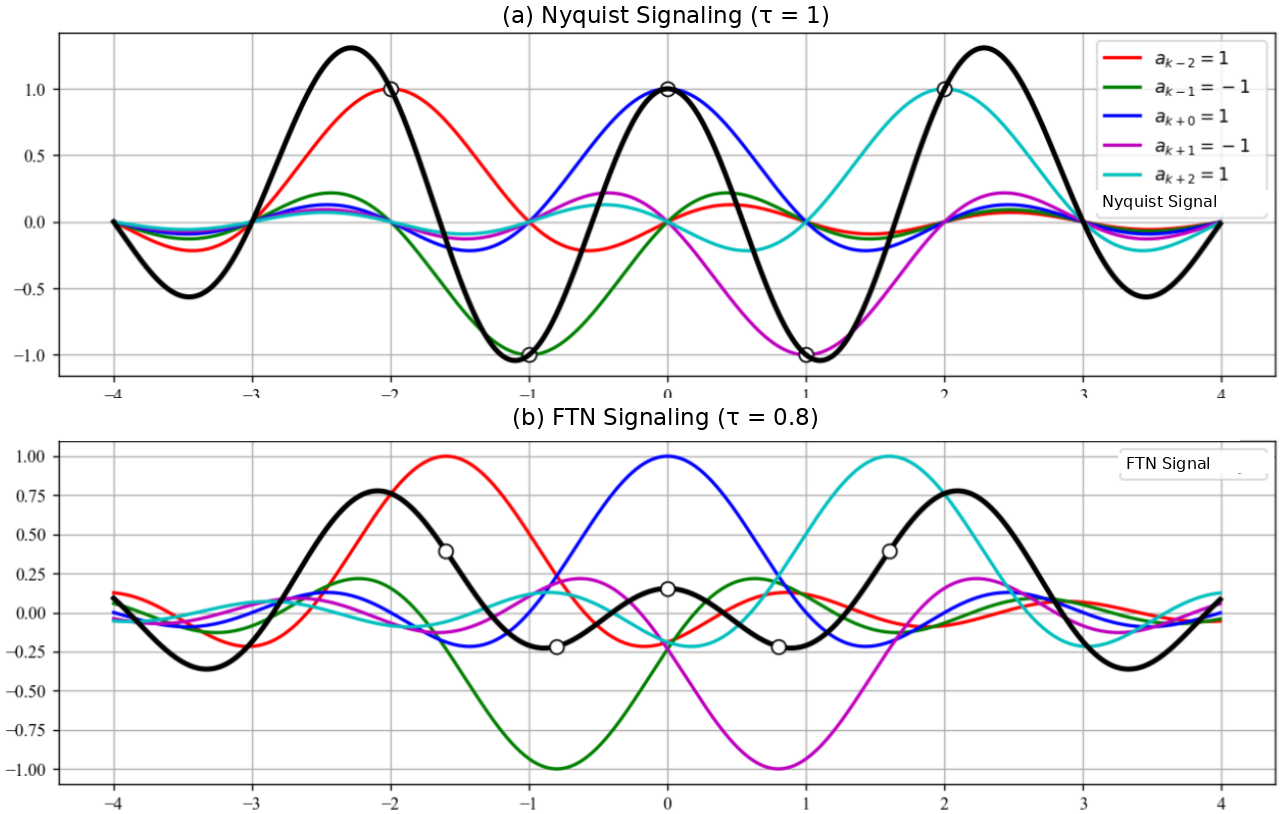}
\caption{Time-domain representation of Nyquist and faster-than-Nyquist (FTN)
signaling obtained using an RRC pulse shape. Colored curves show individual
symbol pulses, the black curve the total transmitted signal, and the circles the
sampling points. (a) Nyquist signaling ($\tau = 1$). (b) FTN signaling
($\tau = 0.8$).}
\label{fig:ftn}
\end{figure}

Fig.~\ref{fig:ftn} shows the FTN transmission structure for a five-symbol BPSK
sequence. As seen in the figure, the detection of a symbol is not independent of
the effect of neighboring symbols. The ISI spread length depends on $\tau$, with
the one-sided ISI length denoted by $N$. In this case, the $k$-th received
sample can be written as
\begin{equation}
\resizebox{0.93\columnwidth}{!}{$
y_k =
\begin{bmatrix} x_N & x_{N-1} & \cdots & x_0 & \cdots & x_{N-1} & x_N \end{bmatrix}
\begin{bmatrix} a_{k-N} \\ \vdots \\ a_k \\ \vdots \\ a_{k+N} \end{bmatrix}
+ w_k.
$}
\end{equation}
For a symbol sequence of length $K$, the system model can be expressed in matrix
form as
\begin{equation}
\mathbf{y} = \mathbf{X}\mathbf{a} + \mathbf{w}.
\end{equation}
Here, $\mathbf{y}$ denotes the received sample vector, $\mathbf{a}$ the
transmitted symbol vector, and $\mathbf{X}$ the Toeplitz-structured interference
matrix composed of the ISI coefficients. Since the ISI is symmetric and spans
only a few taps, $\mathbf{X}$ is a banded Toeplitz matrix, which keeps the
effective detection memory short and motivates a lightweight recurrent model.

%-------------------------------------------------------------------------
%  III. ELMAN-RNN-BASED FTN SIGNAL DETECTION TECHNIQUE
%-------------------------------------------------------------------------
\section{Elman-RNN-Based FTN Signal Detection Technique}

\subsection{Elman RNN and Its Suitability for FTN Detection}
The Elman recurrent neural network (also known as the vanilla or simple RNN) is
the most basic recurrent network structure with a single hidden-state feedback
loop~\cite{elman}. The hidden state at time $t$ is computed as
\begin{equation}
\mathbf{h}_t = \tanh\!\left(\mathbf{W}_x\,y_t + \mathbf{W}_h\,\mathbf{h}_{t-1}
+ \mathbf{b}\right),
\end{equation}
where $\mathbf{W}_x$, $\mathbf{W}_h$, and $\mathbf{b}$ are trainable parameters
and $H$ is the hidden-state dimension. Gated structures such as LSTM and GRU
include additional gate and cell-memory parameters to solve the vanishing-gradient
problem in long-term dependencies. However, in FTN signaling the intersymbol
interference has a short, finite memory limited to a few symbols determined by
$\tau$; hence the long-dependency problem that gating mechanisms address does not
arise in this application. Therefore, the classical Elman RNN is the structure
that models the problem at the correct scale with the fewest parameters and
operations.

\subsection{Bidirectional Single-Pass Architecture}
\begin{figure}[t]
\centering
\includegraphics[width=0.99\columnwidth]{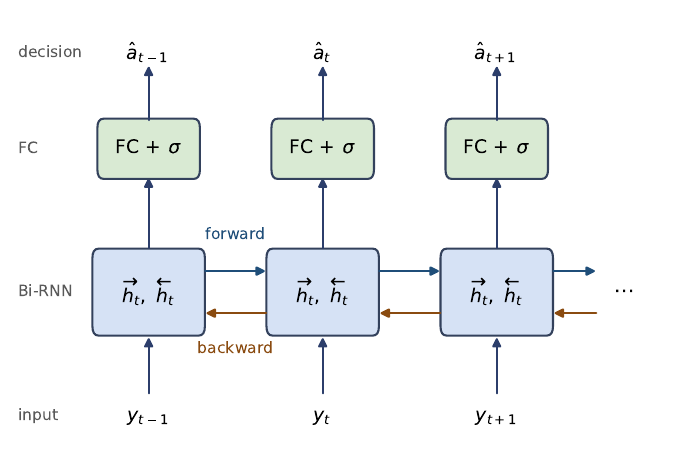}
\caption{Unrolled structure of the proposed bidirectional single-pass Elman RNN
detector. The received sequence is processed in a single pass in the forward
($\protect\overrightarrow{h}_t$) and backward ($\protect\overleftarrow{h}_t$) directions; for
each symbol the hidden states are concatenated and converted into a decision
($\hat a_t$) through a fully connected layer and a sigmoid.}
\label{fig:arch}
\end{figure}

The intersymbol interference channel can be modeled by a finite-state machine,
and BCJR, the optimal symbol-wise maximum a posteriori (MAP) detector, operates
on this trellis with one forward ($\alpha$) and one backward ($\beta$)
recursion~\cite{bcjr}. To mimic this structure, the proposed detector processes
the received sequence $\mathbf{y}=[y_1,\dots,y_L]$ in a \emph{single pass} and
\emph{bidirectionally}:
\begin{align}
\overrightarrow{\mathbf{h}}_t &= \tanh\!\big(\mathbf{W}_x^{f} y_t
+ \mathbf{W}_h^{f}\overrightarrow{\mathbf{h}}_{t-1}+\mathbf{b}^{f}\big),\\
\overleftarrow{\mathbf{h}}_t &= \tanh\!\big(\mathbf{W}_x^{b} y_t
+ \mathbf{W}_h^{b}\overleftarrow{\mathbf{h}}_{t+1}+\mathbf{b}^{b}\big).
\end{align}
The unrolled structure of the proposed architecture is shown in
Fig.~\ref{fig:arch}. The forward and backward hidden states are concatenated for
each symbol and converted into a symbol decision through a fully connected layer:
\begin{equation}
\hat{a}_t = \mathrm{sgn}\!\Big(\sigma\big(\mathbf{w}_o^{\top}
[\overrightarrow{\mathbf{h}}_t;\overleftarrow{\mathbf{h}}_t]+b_o\big)-\tfrac12\Big),
\end{equation}
where $\sigma(\cdot)$ is the sigmoid function. Since the sigmoid is monotonic, at
inference the decision reduces to
$\mathrm{sgn}(\mathbf{w}_o^{\top}[\cdot]+b_o)$; the sigmoid is only required for
the binary cross-entropy loss during training. The forward pass summarizes past interference and the backward pass future
interference; thus the concatenated state $[\overrightarrow{\mathbf{h}}_t;
\overleftarrow{\mathbf{h}}_t]$ of dimension $2H$ acts as a learned
\emph{sufficient statistic} of the BCJR $\alpha/\beta$ metrics. Processing the
whole sequence in a single pass---rather than a separate window per
symbol---makes the per-symbol operation count independent of the window length
and requires neither an explicit trellis nor per-symbol windowing.

\subsection{Training and Computational Complexity}
The network is trained end-to-end with the binary cross-entropy loss and the
NADAM optimizer. The training data are jointly generated from the $7$--$10$~dB
SNR range; early stopping and learning-rate reduction are applied, and evaluation
is performed on a test set completely disjoint from training. Training uses
sequences of length $200$, while complexity is reported over $100$ symbols for
consistency with the M-BCJR reference; since the single-pass per-symbol cost is
independent of the block length, this difference does not affect performance. The number of trainable
parameters of the proposed model, including the separate input and recurrent
bias vectors, is $2H^2+8H+1$, which equals $25$, $43$, and $65$ for $H=2,3,4$,
respectively.

For the single-pass bidirectional structure (the input is scalar), the number of
multiplications and additions over $100$ symbols is $D\,H(H{+}1)\cdot100 +
D\,H\cdot100$ for $D=2$ directions, the number of tanh operations is
$D\,H\cdot100$, and the number of sigmoid operations is $100$; here the
$D\,H(H{+}1)\cdot100$ term corresponds to the recurrent layer and the
$D\,H\cdot100$ term to the per-symbol output (fully connected) layer. The
hardware cost is evaluated using the 10-bit look-up-table (LUT) unit weights
from~\cite{gru} (multiplication: 113, addition: 10, division: 236, exponential:
73, tanh and sigmoid: 1 LUT). Since the nonlinear tanh and sigmoid functions are
implemented with look-up tables, their cost is included in these figures.
Consequently, the advantage of the proposed structure stems from never using the
expensive division (236 LUT) and exponential (73 LUT) operations of the M-BCJR
branch metric as \emph{explicit arithmetic}; the bulk of the M-BCJR cost comes
from these two operations (see Table~\ref{tab:complexity}).

%-------------------------------------------------------------------------
%  IV. SIMULATION RESULTS
%-------------------------------------------------------------------------
\section{Simulation Results}

\begin{figure}[t]
\centering
\includegraphics[width=0.99\columnwidth]{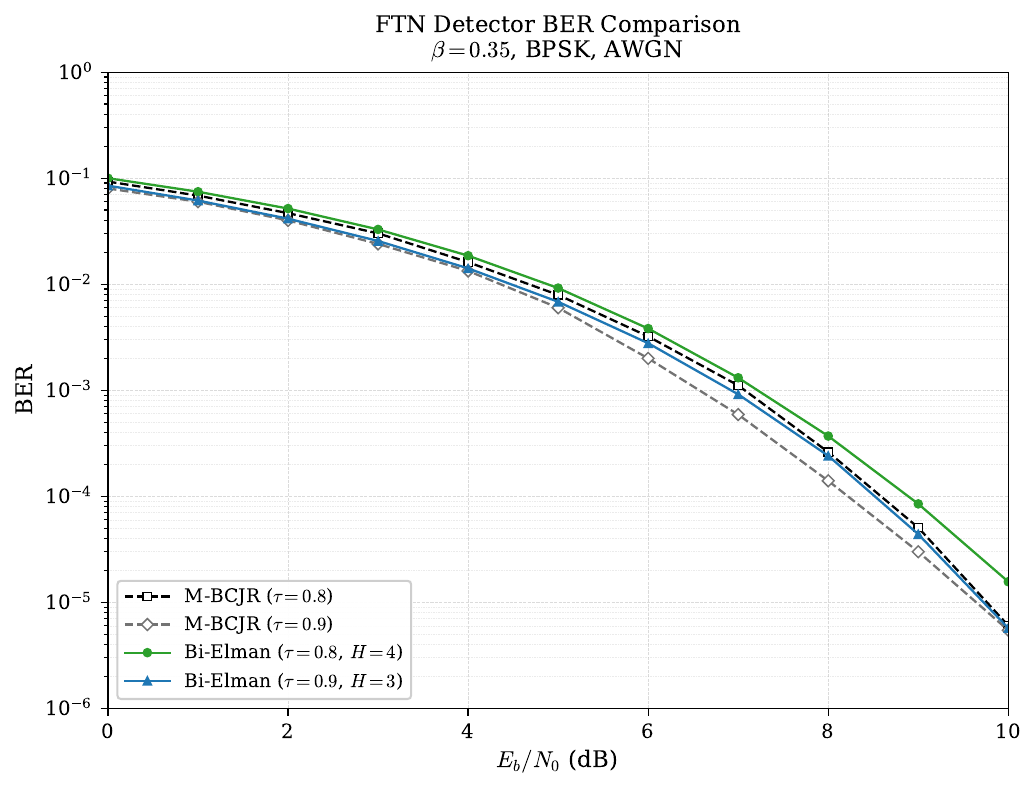}
\caption{BER comparison of the proposed bidirectional single-pass Elman RNN and
M-BCJR for $\tau=0.8$ and $\tau=0.9$ (BPSK, AWGN).}
\label{fig:ber}
\end{figure}

The proposed detector is evaluated under a root-raised-cosine pulse with roll-off
factor $\beta=0.35$ and an AWGN channel for $\tau=0.8$ and $\tau=0.9$. The
near-optimal M-BCJR algorithm is used as reference. The system and training
parameters are given in Table~\ref{tab:sys}, the LUT-based complexity comparison
in Table~\ref{tab:complexity}, and the BER performance in Fig.~\ref{fig:ber}. The
BER results at selected SNR values are additionally listed numerically in
Table~\ref{tab:ber}.

As seen in Fig.~\ref{fig:ber}, the proposed bidirectional single-pass Elman RNN
closely follows the M-BCJR curve for both compression factors, providing a BER
within about $0.1$--$0.4$~dB of the optimum. This performance is achieved with
38\% lower LUT cost for $\tau=0.9$ ($H=3$) and 46\% lower for $\tau=0.8$ ($H=4$).
The more aggressive $H=2$ configuration offers 67\% lower cost for $\tau=0.9$,
so $H$ can be selected according to the performance--cost budget of the target
receiver. These results show that a classical Elman RNN can attain performance
close to M-BCJR in FTN detection with very few parameters ($25$--$65$) at
markedly lower hardware cost. The same behavior holds for both compression
factors across the entire $0$--$10$~dB range, confirming a consistent gain. The
sub-decibel performance gap together with the absence of explicit
division/exponential operations (shorter critical path, higher numerical
stability) makes the proposed detector a practical M-BCJR alternative for
real-time receivers.

Using the same LUT methodology, the fixed-kernel CNN detector~\cite{cnn} reports
about a 46\% reduction for BPSK and $\tau=0.9$; the proposed Elman structure
achieves a comparable reduction (38--67\%) gating-free and with far fewer
parameters, and compared with GRU-based detectors~\cite{gru} the absence of gate
structures further reduces the operation count. A direct comparison in an
identical environment with these lines is left to future work.

\begin{table}[t]
\centering
\caption{\textsc{BER Results at Selected SNR Values}}
\label{tab:ber}
\setlength{\tabcolsep}{4pt}
\begin{tabular}{c cc cc}
\toprule
 & \multicolumn{2}{c}{$\tau=0.8$} & \multicolumn{2}{c}{$\tau=0.9$} \\
\cmidrule(lr){2-3}\cmidrule(lr){4-5}
$E_b/N_0$ & M-BCJR & Bi-Elman & M-BCJR & Bi-Elman \\
(dB) & & ($H{=}4$) & & ($H{=}3$) \\
\midrule
4  & $1.6\times10^{-2}$ & $1.9\times10^{-2}$ & $1.3\times10^{-2}$ & $1.4\times10^{-2}$ \\
6  & $3.2\times10^{-3}$ & $3.8\times10^{-3}$ & $2.0\times10^{-3}$ & $2.8\times10^{-3}$ \\
8  & $2.6\times10^{-4}$ & $3.7\times10^{-4}$ & $1.4\times10^{-4}$ & $2.4\times10^{-4}$ \\
10 & $6.0\times10^{-6}$ & $1.6\times10^{-5}$ & $5.5\times10^{-6}$ & $5.7\times10^{-6}$ \\
\bottomrule
\end{tabular}
\end{table}

\begin{table}[t]
\centering
\caption{\textsc{System and Training Parameters}}
\label{tab:sys}
\begin{tabular}{l l}
\toprule
Parameter & Value \\
\midrule
Modulation & BPSK \\
Pulse shape & RRC, roll-off $0.35$ \\
Compression factor & $0.8$ and $0.9$ \\
Channel & AWGN \\
Detector & Bidirectional single-pass Elman RNN \\
Hidden size $H$ & $2$--$4$ \\
Number of parameters & $25$--$65$ \\
Optimizer & NADAM \\
Loss & Binary cross-entropy \\
Sequence length & $200$ symbols \\
\bottomrule
\end{tabular}
\end{table}

\begin{table}[t]
\centering
\caption{\textsc{LUT-Based Complexity Comparison (100 Symbols)}}
\label{tab:complexity}
\begin{tabular}{l c c c}
\toprule
Method & Division/Exp & LUT & Ratio \\
\midrule
M-BCJR ($\tau{=}0.9$)            & yes & $594\,750$    & $1.00$ \\
Bi-Elman ($\tau{=}0.9$, $H{=}3$) & no  & $369\,700$    & $0.62$ \\
Bi-Elman ($\tau{=}0.9$, $H{=}2$) & no  & $197\,300$    & $0.33$ \\
\midrule
M-BCJR ($\tau{=}0.8$)            & yes & $1\,095\,896$ & $1.00$ \\
Bi-Elman ($\tau{=}0.8$, $H{=}4$) & no  & $591\,300$    & $0.54$ \\
\bottomrule
\end{tabular}
\end{table}

%-------------------------------------------------------------------------
%  V. CONCLUSION
%-------------------------------------------------------------------------
\section{Conclusion}
In this study, a bidirectional, single-pass Elman RNN detector was proposed for
FTN-BPSK signal detection. By mimicking the forward-backward recursion structure
of BCJR with a very small recurrent network, the proposed structure attained a
BER close to M-BCJR (within about $0.1$--$0.4$~dB of the optimum) for $\tau=0.8$
and $\tau=0.9$, at 38--46\% lower LUT-based hardware cost and without using any
explicit division/exponential arithmetic operations; the more compact $H=2$
configuration reduces the cost by up to 67\% for $\tau=0.9$. To the best of our
knowledge, this is the first study to investigate the applicability of the
classical Elman RNN architecture to the FTN detection problem.

The results reveal
that a small recurrent network chosen appropriately for the nature of the problem
can offer near-optimal performance without much more complex architectures,
making it a cost-effective candidate for receivers with limited hardware
resources. Future work aims to extend the method to QPSK and lower $\tau$ values
and to further reduce complexity through weight quantization.

%-------------------------------------------------------------------------
%  ACKNOWLEDGMENT  (sağ kolona taşı: referansların hemen üstünde)
%-------------------------------------------------------------------------
\newpage
\section*{Acknowledgment}
This work was supported by The Scientific and Technological Research Council of
T\"urkiye (T\"UB\.ITAK) under Project No.~122E236.

%-------------------------------------------------------------------------
%  REFERENCES
%-------------------------------------------------------------------------

\end{document}